\documentstyle[12pt, aaspp]{article}
\begin{document}
 
\def\today{\number\year\space \ifcase\month\or  January\or February\or
        March\or April\or May\or June\or July\or August\or
September\or
        October\or November\or December\fi\space \number\day}
\def\fraction#1/#2{\leavevmode\kern.1em
 \raise.5ex\hbox{\the\scriptfont0 #1}\kern-.1em
 /\kern-.15em\lower.25ex\hbox{\the\scriptfont0 #2}}
\def\spose#1{\hbox to 0pt{#1\hss}}
\def\simlt{\mathrel{\spose{\lower 3pt\hbox{$\mathchar''218$}}
     \raise 2.0pt\hbox{$\mathchar''13C$}}}
\def\simgt{\mathrel{\spose{\lower 3pt\hbox{$\mathchar''218$}}
     \raise 2.0pt\hbox{$\mathchar''13E$}}}
\def\etal{et al. }
 
\title{Halting Planetary Migration}
\author{M. Lecar \& D. D. Sasselov\altaffilmark{1}}
\affil{Harvard-Smithsonian Center for Astrophysics, 60 Garden St., Cambridge MA 
02138}
\altaffiltext{1}{Alfred P. Sloan Foundation Fellow}
 
\begin{abstract}
When Jupiter's Roche Lobe radius exceeded the scale height of the
protoplanetary disk, Jupiter opened a gap in the disk. When the gap was
wide enough, tidal torques from the disk interior and exterior to Jupiter
were suppressed and migration continued on the accretion time
scale. In the 'minimum solar nebula' about two Jupiter masses of gas
remained between Jupiter and Saturn and about five Jupiter masses
between Jupiter and Uranus. Unless all but a Jupiter mass of the outer
disk was removed, Jupiter would have migrated into the Sun on the
accretion time scale. So far, no mechanism has been postulated to remove
the outer disk except for photo-evaporation which might work exterior
to Saturn. For extra-solar Jupiters at an AU or less from the central
star, photo-evaporation would not remove the outer disk.
 
We propose a new mechanism, relying on the irradiation of the exposed
edge of the gap by the Sun, to cause gas from the outer disk to cross the
gap and become part of the inner disk, whence it was later accreted by
the Sun.

\end{abstract}

\section{Introduction}
Current theories of disk dispersal remove the 'inner disk' (the disk
between the Sun and Jupiter) by accretion, and remove the disk exterior
to Saturn by photo-evaporation (see Hollenbach, Yorke, \& Johnstone 2000). 
However, even in the minimum solar nebula proposed by Hayashi (1981),
about 2 Jupiter masses of material remain between Jupiter
and Saturn.
After Jupiter carves a gap in the disk, the inner disk accretes onto the
Sun, and with more than a Jupiter mass exterior to Jupiter, Jupiter
follows the inner disk into the Sun. Two detailed numerical simulations
(Nelson \& Benz 1999; Nelson et al. 1999) reproduce the creation of the
gap, but do not follow the further migration of the planet on the
accretion time scale.

Even before these simulations, we concluded that to keep Jupiter from
migrating into the Sun (on the accretion time scale) we would have to
get rid of the disk exterior to Jupiter (the outer disk). We came up
with a number of fanciful speculations but were unable to explain why
they were not picked up by the simulations. We were driven to read our
own recent work on 'The Snow Line' (Sasselov \& Lecar 2000, thereafter
SL2000)
which did contain an effect not
included in the above simulations, i.e., the irradiation of the disk by the 
Sun. In a disk with negligible accretion ($< 10^{-8} M_{\odot} {\rm yr}^{-1}$) 
typical
of old T Tauri disks, this irradiation controls the temperature of the disk.

In this paper we suggest that the irradiation of the parent star can control
the rate of migration and even halt it altogether in the inner regions 
($< 10~AU$) of a typical protoplanetary disk.

\section{The Irradiation Model}
Our model is that of a star surrounded by a flared disk,
the same as in SL2000.
Our disk has a surface gas mass density
$\Sigma$= $r^{-3/2}{\Sigma}_0$, with $r$, the distance from the star (or Sun),
in AU and 
${\Sigma}_0$= 10$^3$g~cm$^{-2}$, which is a standard minimum-mass solar
nebula model. The dust and gas are assumed well mixed. The emergent
spectrum of the star is calculated with a stellar model atmosphere code
with Kurucz (1992) line lists and opacities. The disk intercepts the
stellar radiation $F_{irr}(r)$ at a small grazing angle 
(typically 3$^o$ at 1~AU). After the planet opens a gap,
the {\em wall} at the gap's outer rim is exposed to the star. In this
paper we work with a fairly narrow gap (see \S 3.), which exposes $\sim 1/5$
of the wall to the direct starlight. We do this because our mechanism (\S 3.)
starts working as soon as even a partial gap has been carved out; the much
wider gaps seen in simulations by, $e.g.$, Nelson \& Benz (1999) are a later
phase in the evolution of the disk. We assume the gap wall to be 
inclined at 60$^o$ to the disk plane at this early stage in the development
of the gap (a range of 45$^o$-90$^o$ does not make a difference). 
Thus, the flux intercepted by a unit
surface area of the gap's wall can be 5 or more times
greater than in the unperturbed disk. In our computation we account for the
cooling taking place in the midplane of the gap's outer wall which is in the
shadow of the inner disk and the pressure gradient which develops.

We compute the radiative transfer for the gap's outer rim in
the same fashion as for the rest of the disk (for details, see
SL2000). In particular, we use dust grains with
properties which best describe the disks of T Tauri stars.
Our calculation for the gap's wall is generally valid for
$r \geq 0.2$~AU, so that the temperatures never exceed 1500-1800K
and we do not consider dust sublimation; the dust is present at all
times and is the dominant opacity source. This is required for our
radiative transfer solution.

The midplane temperature in our passive unperturbed disk is derived from the 
requirements
of hydrostatic and radiative equilibrium, and the balance between
heating by irradiation and radiative cooling, $\sigma T^4(r) = F_{irr}(r)$.
Our midplane temperature scales
as $T(r)$= $T_0 r^{-3/7}$~K. The scaling coefficient is $T_0 = 140$;
the actual numerical solution for $T(r)$ differs 
insignificantly from this simple form in the range of interest to
us ($0.1<r<10AU$). This result is very insensitive to the surface density,
${\Sigma}_0$; we find that the changes in
$T(r)$ are within $\pm 10K$ for an increase or decrease
in $\Sigma_0$
of a factor of 10 (SL2000).

From these calculations here we find, for a gap which is twice the scale
height, that the effective photosphere of
the gap's exposed wall is typically about 1.6 times hotter than the 
corresponding
gas temperature at that distance in the unperturbed disk.
The ratio is indeed almost constant, slowly varying with $r$
from 0.2 to 5.0~AU. The midplane
temperature of the disk returns to the unperturbed temperature
within only about 2 gas scale heights behind the gap wall, also a slowly
varying function of $r$. A large pressure gradient develops between the
heated part of the wall and the midplane section which is in the shadow.

We note that in  a disk with a high accretion rate, irradiation plays
no role in its thermal structure even close to the star (see D'Alessio
et al. (1999) for such examples). However, planet forming disks may
not have such high accretion rates ($> 10^{-8} M_{\odot} {\rm yr}^{-1}$),
as evidenced by the large number of passive disks around old T Tauri stars.
As in SL2000, we postulate that the latter disks comprise
the basis (parameters, dust properties, etc.) of our unperturbed
protoplanetary disk model.

\section{The Mechanism}
We emphasize that the unperturbed disk is razor thin and only slightly
flared. The opening angle is $h/r=c/r\Omega = 0.044 r^{2/7}$, where 
$h$ is the scale height of the disk,
$\Omega$ is the angular frequency ($r\Omega$ is the circular
orbital velocity). For the unperturbed disk, before a gap opens,
$T(r)$= $140 r^{-3/7}$~K, so the sound speed is $c=1.32$~km.s$^{-1} r^{-3/14}$.
At Jupiter's distance, $r_{\rm J}= 5.203$, $c=0.821$~km.s$^{-1}$ and
$h/r=0.0705$. Because the disk is so thin, only a small fraction of
the solar radiation is intercepted and absorbed. This fraction is very
sensitive to small perturbations in the geometry; hills expose more area
and heat up, while valleys, in the shadow, cool down.
 
We compare this disk with the disk after Jupiter has begun to carve a gap.
We focus on an initial gap of radial extent $h$ on either side of Jupiter.
The inner disk has $r<r_{\rm J}-h$ and the outer disk has $r>r_{\rm J}+h$.
The gas in the gap is in the shadow of the inner disk and cools down. 
Part of the outer disk, formerly in the
shadow of the inner disk, is now exposed. We call this the 'wall'. The
height of the wall is:
$$
\Delta h = h(r_{\rm J}+ h) - h(r_{\rm J}-h) = 2h{{dh}\over {dr}},
$$
where
$$
{{\Delta h}\over {r}} = 0.114r^{{2}\over {7}}{{h}\over{r}} = 0.182 {{h}\over{r}}
$$
at $r_{\rm J}$.
The calculation described in \S 2 is now applied to this wall and 
determines that the temperature at the wall increases
by a factor of 1.6, and that this higher temperature extends for a
distance $\sim 2h$ behind the wall.
 
The 'wall' increases the area directly exposed to sunlight.
The Sun's luminosity, $L_{\odot}$, can be expressed in units of
the orbital kinetic energy of Jupiter 
($E_{\rm J}={{1}\over{2}}M_{\rm J}v_{\rm J}^2 = 1.6\times 10^{42}$ ergs)
and the orbital period of Jupiter ($T_{\rm J}$=11.9~years). In those units,
$L_{\odot}= 3.826\times 10^{33}$~ergs.s$^{-1}$= 0.9$E_{\rm J}/T_{\rm J}$.
The luminosity intercepted by the
wall is $L_{\rm int} = 0.013 L_{\odot}$, so enough energy is intercepted 
to move a Jupiter mass of gas an appreciable distance
in about 80 Jupiter periods.

The mass transfer across the wall is:
$$
{{dm}\over {dt}} = 2{\pi}{\Sigma}rc = 1.20\times 10^{29} g/yr
= 0.06 M_{\rm J}/yr = 0.75 M_{\rm J}/T_{\rm J},
$$
where $M_{\rm J}$ is the Jupiter mass.
At the sound speed, the time to cross the gap,
$tc = 2h/c \cong 3~yrs \approx {{1}\over{4}}T_{\rm J}$.
The energy transferred across the wall is
$dE/dt = {{1}\over{2}}c^2{dm/dt} = 0.013M_{\rm J}/T_{\rm J}\approx L_{\rm int}$.

Looking at the dynamics in somewhat more detail, in the unperturbed disk, on
a hydrodynamic time-scale (i.e., neglecting accretion),
$$
{{dv}\over {dt}} = -{{GM_{\odot}}\over {r^2}} + r{\Omega}^2 + 
(-{{1}\over {\rho}} {{dP}\over {dr}}) = 0 ,
$$
where $\rho$ is the gas density and $P$ is the pressure. 
The zeroth order ($r=r_0$, ${{dr}\over {dt}} =0$)
gives us, for the acceleration due to the pressure 
gradient:
$$
g \equiv -{{1}\over {\rho}} {{dP}\over {dr}}\cong {{c^2}\over {r}}.
$$
A calculation yields the numerical factor 1.07. 
The ratio of $g$
to $r{\Omega}^2$ is $(c/r\Omega)^2 =(h/r)^2$, so this term can be ignored.
The first order gives us:
$$
g \cong {{c^2}\over {h}} \cong c{\Omega}_0 \cong 
({{\Delta T}\over {T}})^{{1}\over {2}}c_0{\Omega}_0,
$$
where we have $r=r_0(1+x)$ and $\ddot{x} = -{\Omega}_0^2x + g$, and
from \S~2 we have ${{\Delta}T}/T =1.6$.

After the 'wall' is exposed:
$$
g = {{c^2}\over{h}} = c{\Omega},
$$
which is an increase by a factor of $r/h$, and in addition
$c$ increases by $(1.6)^{1/2}$.
At this point, after the gap, $g = c{\Omega}$ is comparable to the 'restoring
force' at the maximum extension, $er$:
$$
er{\Omega}^2 = eV{\Omega} = c{\Omega}.
$$

Note that in this process we conserve angular momentum. Denoting the
angular momentum  per unit mass by $L$:
$$
L^2 = GMa(1-e^2) = GMq(1+e),
$$
where $q=a(1-e)$ is the perihelion distance and $e$ is the eccentricity.
$\em Conserving$ angular momentum, $L=const.$, we find $dq/q= -de/(1+e)
\approx -e$, while the increase in 
the semi-major axis is insignificant, $da/a= d(e^2)/(1-e^2) \sim 0$,
for small $e$.

At this point we have described a mechanism by which gas can cross the gap
whence it will continue to accrete onto the central star. The gas which
crossed the gap, on a sound-speed time scale, will have to be replenished
on an accretional time scale. But, where without this mechanism, Jupiter
would have been driven into the Sun by the inwardly moving outer disk, we
now allow a substantial fraction of the outer disk to cross the
gap into the inner disk.
 
We further suggest, admittedly in a more speculative vein, a mechanism
that might speed up the replenishment. Previous work (Lin \& Papaloizou
1980, Goldreich \& Tremaine 1980; and more recently, Bryden et al. 1999, 
Kley 1999) has shown that
Jupiter exerts tidal torques on it's outer disk and induces trailing
density enhancements in 
a tightly wrapped spiral. This series of ridges (which intercept
sunlight directly) and valleys (which cool down because they lie in the
shadow of the ridges) initiates our mechanism at every ridge. In effect,
they act like a closely spaced grating of mini-gaps and our mechanism
will transfer material across each gap. We re-emphasize the fact that
the intercepted sunlight depends on $h/r$ which is a slowly increasing
function of $r$ (increasing as $r^{2/7}$). The intercepted flux does $not$ fall
off as $r^{-2}$, because the area of the intercepting annulus increases as
$r^2$.

This is as far as these estimates should be taken. Further details should
be determined from a numerical simulation. We suggest that such a
simulation could be performed using a non-axisymmetric thin disk model,
including Jupiter, but with the solar heat input at the wall included.
All the previous simulations, that we are aware of, were 
accretionally heated, and so ignored this effect. We predict that a
simulation including this effect will show a natural stop to the
migration at substantial distances from the sun.

\section{Summary}
Unless the outer disk (e.g., beyond Jupiter) is removed,
migration on the accretion time scale
cannot be halted by any mechanism so far discussed in the literature.
We propose a mechanism to halt migration in the inner regions of
protoplanetary disks in which irradiation by the central star dominates
over accretion. Once a gap is partially open, stellar irradiation
provides sufficient energy to the outer wall of the gap to move material
across the gap. The mechanism acts as a semi-permeable
membrane: material moved outwards returns to the rim of the gap,
but material moved inwards, across the gap, never returns.
Any planet which opens a gap should trigger the mechanism.

\medskip
\acknowledgments{We are grateful to John Papaloizou for useful discussions, 
and to the referee for thoughtful and helpful
comments. DDS acknowledges support from the Alfred P. Sloan
Foundation.}

\end{document}